# *In silico* drug repositioning for COVID-19 using absolute binding free energy calculations

Théau Debroise[1], Rose Hoste[1], Quentin Chamayou[1], Hervé Minoux[2], Bruno Filoche-Rommé[2], Marc Bianciotto[2], Jean-Philippe Rameau[2], Laurent Schio[2], Maximilien Levesque[1]

[1] Aqemia, Paris, France

[2] Molecular Design Sciences - Integrated Drug Discovery, Sanofi R&D, Vitry-sur-Seine, France

*Aug 10th, 2021*

## Abstract

*Since the rise of the SARS-CoV-2 pandemic in the winter of 2019, the need for an affordable and efficient drug has not yet been met. Leveraging its unique, fast and precise binding free energy prediction technology, Aqemia screened and ranked FDA-approved molecules against the 3ClPro protein. This protease is key to the post-translational modification of two polyproteins produced by the viral genome. We propose in our top 10 predicted molecules some drugs or prodrugs that could be repurposed and used in the treatment of COVID cases.*

## Introduction

The worldwide spread of severe acute respiratory syndrome coronavirus 2 (SARS-CoV-2) occurred over the course of a few weeks. The first reports of cases in the province of Wuhan, China, started in December 2019 and the World Health Organization declared it a pandemic only four months later. As of December 2020, more than 202M people have been infected and 4.29M died.[1] Given its scale, its rapid spread and high mortality rate, there is an urgent need to identify small molecule drugs (SMD) that can cure contaminated people with severe cases. It is important to note that even once vaccines are available and abate the pandemic, SMDs are still relevant for the people that cannot avoid contamination: no vaccine has 100% efficiency and it is impossible to vaccinate 100% of the world population.

As essential it is to find a drug against the coronavirus-disease 2019 (Covid-19), it remains a lengthy, costly and possibly unsuccessful process. The average cost of bringing a new compound to the market nearly doubled between 2003 and 2013 to $2.6 billion, according to the Tufts Center for the Study of Drug Development.[2] These same challenges have increased the lab-to-market timeline to 12 years, with 90 percent of drugs washing out in one of the phases of clinical trials.[2] Even if those numbers are discussed in their values, the orders of magnitude are well agreed upon and the standard drug discovery workflow is inadequate to face this challenge in an accelerated time frame.

Alternatively, a drug repositioning strategy[3,4] can be used to accelerate the process by identifying approved drugs as well as clinically tested drug candidates which have the potential to treat COVID-19. In general, the drugs already in use were deemed safe in prior studies and have an established large-scale industrial production. Finding one of these molecules with antiviral activity against coronavirus would be a solution to rapidly deal with the emergency of COVID-19.

A large scale screening approach leads to a more exhaustive search and possibly, unexpected discoveries compared to a more classical approach. These screenings of molecules that are already approved can be applied both in wet-lab (*in vitro*) or virtually (*in silico*) for evaluating the ability of known drugs to bind against the essential proteins of SARS-CoV-2 and hence identify those with potential antiviral activity. *In vitro* screenings ask for the use of either the purified target protein or cells infected with the SARS-CoV-2. On one hand, the purified protein is easy enough to manipulate, but it is only a *proxy* to what happens in the virus in the much more complex media that is the human body. On the other hand, live viruses are complex to manipulate and require specific labs and security measures. Due to this limitation and the lower cost associated with it, *in silico* screenings have gathered the attention of the scientific community.

However, most virtual screening suffers from inaccurate protein-ligand affinity predictions. A common source of the prediction error comes from fast empirical scoring functions used for docking. Physics-based scoring functions should be more accurate but they become expensive when entropic contributions are fully taken into account like in absolute binding free energy calculations (ABFE).

To bypass this trade-off between accuracy and speed, we have leveraged our fast and precise affinity predictor to screen and rank a library of 1400 FDA-approved drugs in only a few minutes. Our predictor is a free energy calculation which can be expressed as follow :

$$\Delta G_{calc} = \Delta E_{MM} + \Delta G_{solv} \quad \text{(equation 1)}$$

$$\Delta G_{solv} = \Delta G_{solv}(PL) - \Delta G_{solv}(P) - \Delta G_{solv}(L) \quad \text{(equation 2)}$$

where $\Delta E_{MM}$ is a Molecular Mechanic (MM) calculation summing electrostatic and Van der Waals contributions to protein-ligand interaction and $\Delta G_{solv}$ is the difference of solvation energy between the complex (PL), the protein (P) and the ligand (L). Each of these solvation calculations is based on the molecular density functional theory (MDFT)[5].

## Target Selection

The SARS-CoV-2 genome is made up of 30,000 nucleotides, coding for 28 proteins.[6] Of these 28 proteins, 16 non-structural proteins come from two large polyproteins, pp1a and pp1b. These proteins, responsible for example of the shutdown of the cell defences against viruses[7], are then cleaved by two viral proteases. These two proteins are thus responsible for the maturation of the majority of the machinery of

SARS-CoV-2. The first one, papain-like protease (PLPro) is a cysteine protease with a zinc finger domain.[8] The second one, known as the main protease (Mpro)[9,10], also called 3CLPro is a key protein in the activity of viral replication as it is responsible for eleven of the sixteen proteolytic cleavages.[11] Its predominant role in the virus life cycle and the lack of closely related homologues in humans[12] point out 3CLPro as an attractive target for specific inhibitor design with very less chance of off-targeting.

The targeting of viral proteases is a well-known and efficient strategy, which has even been employed in the treatment of HIV and hepatitis.[13] As it stands, the inhibition of 3CLPro should stop the replication of the virus with very little chance of off-targeting, as it has been shown on the main protease of closely related viruses SARS-CoV-1 and MERS-CoV[12,14,15]. This cysteine protease works by stabilizing the thiolate state of a cysteine residue *via* a network of hydrogen bonds. The thiolate is then free to react with a peptidic amide bond. The oxyanion formed following the attack by the thiolate on the carbonyl is stabilized by three hydrogen bonds with Gly 143, Ser 144 and the backbone of Cys 145. This eventually results in the cleavage of the peptide bond (see annex 3).[16]

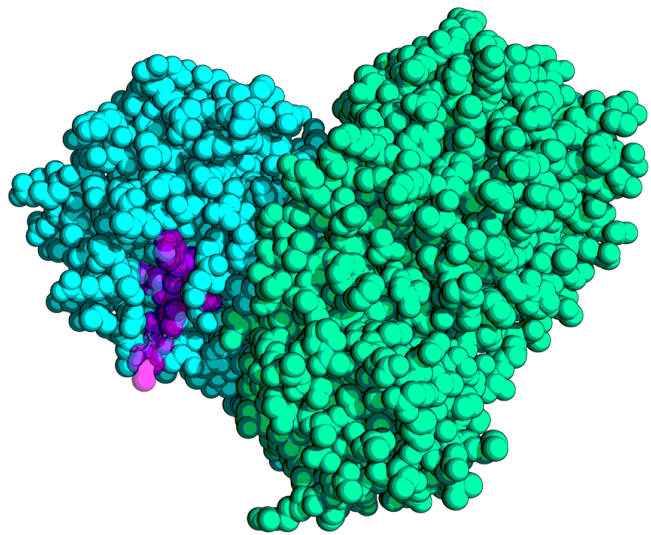

*Figure 1: Crystal structure of the 3CLPro protease of SARS-CoV-2 co-crystallized with the inhibitor N3.(PDB ID: 6LU7)[2]. Protomers A and B are in cyan and green, respectively, and the inhibitor is shown in magenta.*

Research groups sought out to exploit the large homology between the 3CLPro of SARS-CoV-1 and SARS-CoV-2 - with more than 96% matching sequence - to accelerate the design of protease inhibitors by transferring knowledge they already had of the 2003 SARS-CoV-1 virus.[11] The catalytic site of both proteins is situated in the groove formed between the first and second domain of the protein, the third and final one being essential for the formation of a homodimer. The twelve mutations between the SARS-CoV-1 and SARS-CoV-2 protease only influence the affinity of the protease for itself, leading to a more stable homodimer in the case of SARS-CoV-2 and subsequently, a better activity.[17] Usual drugs for the SARS-CoV-1 main protease are covalent inhibitors which take advantage of residue reactivity on the main cysteine to anchor themselves. These inhibitors are usually made up of two main sub-structures : a warhead, that will chemically react with the cysteine of the catalytic dyad, and a recognition site that will interact with the sub-pockets prior to the chemical reaction.[18]

In Figure 2, sub-pockets of the binding site are depicted. They are conserved between coronaviruses[19] and each pocket is used to recognize an amino acid of the peptidic substrate.[20] Sub-pocket 2 (S2) and sub-pocket 4 (S4) both recognize small hydrophobic residues, such as leucine, valine and alanine. Sub-pocket 3 (S3) is less specific in the residue it can accommodate, which ranges from lysine to methionine. Sub-pocket 1 (S1) is, in contrast, highly specific as it only recognizes glutamine residue. There is no known protease in the human proteome with the same specificity.[21–23] A molecule that binds to these sub-pockets with a strong affinity might be a good candidate for further investigations in the treatment of COVID-19.

In addition to the elucidation of the mechanism of the hydrolysis of the peptide bond, several research groups have done analyses of the energetic contribution of specific residues. These analyses broke down the interaction of an inhibitor or substrate in a per-residue basis.[16,24–26] Statistical analyses over a large dataset of crystallographic structures are also reported.[27] These studies showed that a few key residues, like Asn 142, Glu 166, Met 165, Glu 189 and Thr 26 also have a big impact on the affinity of the inhibitor while not participating in the proteolytic mechanism. These residues take part in all subpockets of the active site, showing that an inhibitor should extend throughout all sub-pockets to have the best binding affinity possible.

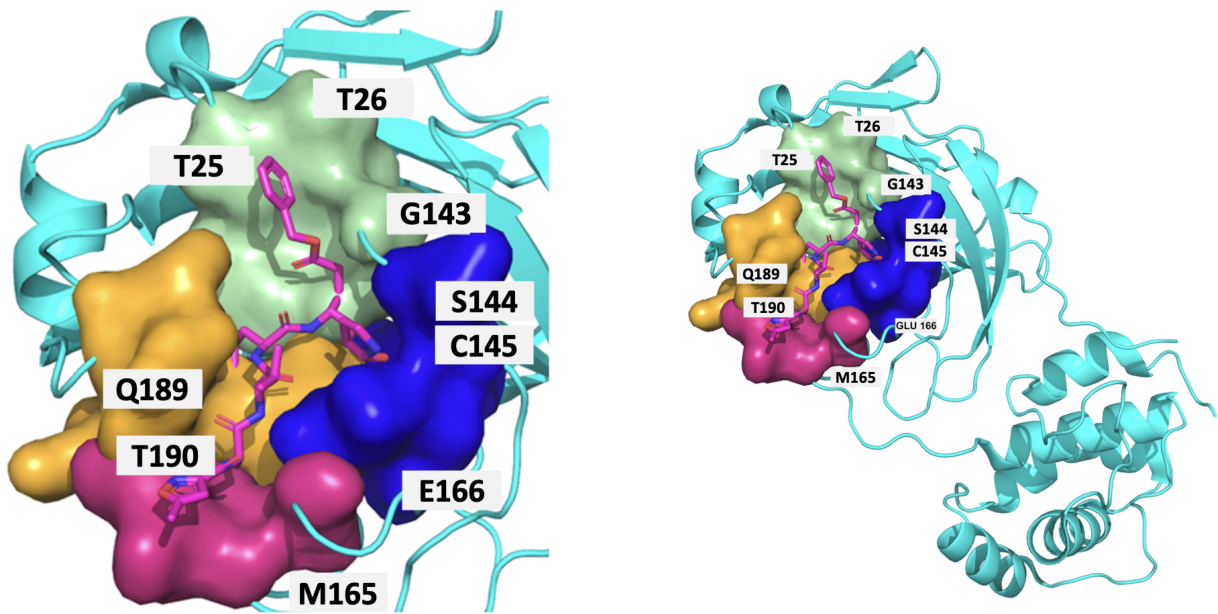

*Figure 2: SARS-CoV-2 protein: 3-Chymotrypsin like protease (3CLPro, PDB ID 6LU7) binding site illustration, with ligand N3 (magenta). S1 (blue), S2 (orange), S4 (magenta), S1' (pea) subpockets.*

The majority of publicly known inhibitors of both SARS-CoV-1 and SARS-CoV-2 main protease are covalent inhibitors that react with the catalytic cysteine. The innate reactivity of the warhead structures leads to potential off-targeting by reacting with other proteins if there are no specific or well-oriented interactions between the drug and the target protein.[28] *In vitro* screenings have been able to identify known drugs as potential 3CLPro inhibitors.[29–31] These screenings revealed the possible repurposing of non-covalent inhibitors such as Masitinib, Vilazodone, Lapatinib or Imatinib.

Following the release of the first crystallographic structures in March 2020, *in silico* (or virtual) screening has been widely used to identify SARS-CoV-2 proteins inhibitors (> 100 publications in 2020 on PubMed). Most of the studies use empirical scoring functions from docking softwares. Screening using MM/GBSA-PBSA physics-based scoring functions also puts forward some interesting known drugs like dipyridamole or HIV-anti protease inhibitors[32,33]. Among them, dipyridamole has been experimentally proven active[34].

## Target Preparation

The affinity prediction methodology developed and applied to obtain the following results is based on the scoring of protein-ligand complexes. Before scoring such complexes, we need to generate a bound pose of the ligand inside the protein structure. Rigid docking is widely deployed to yield such a complex without taking into account the flexibility of the receptor. In order to do so, we decided to use an ensemble docking strategy. Ensemble docking is now well established in the field of early-stage drug discovery and allows one to dock each ligand against multiple rigid receptor conformations. By doing ensemble docking we have a better chance to predict an accurate ligand pose.

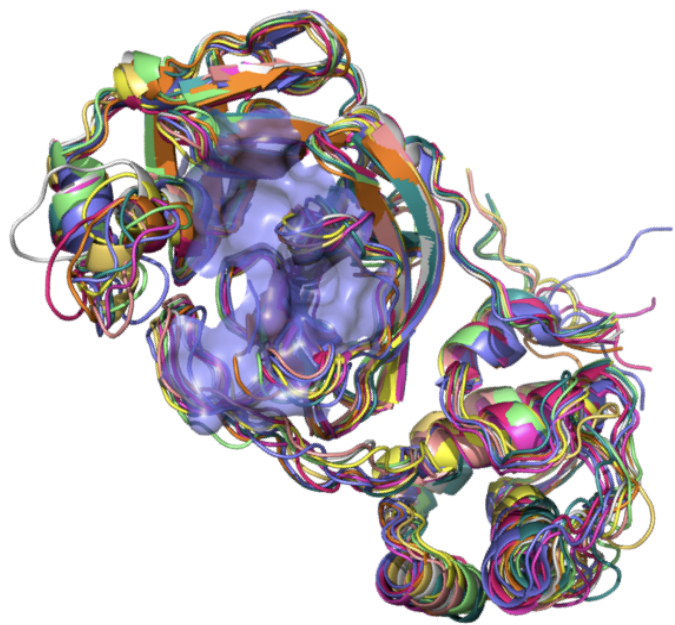

*Figure 3: Structural alignment of all cluster centroids aligned on 6LU7. Centroids are extracted from the 6Y84 molecular dynamics trajectory. The 10 structures are the centroïd of the 10 RMSD clusters covering over 90% of the trajectory. The binding site shown one the surface (pale purple) is from 6LU7 for easier visualization.*

The "ensemble" of receptor conformations can be obtained by using molecular dynamics simulations. On March 27th, 2020 the D. E. Shaw Research Group released a 100 µs molecular dynamics simulation[35] starting from the apoenzyme structure and was determined by X-ray crystallography (PDB ID: 6Y84). We have clustered this trajectory with GROMOS RMSD-based cluster algorithm[36]. The RMSD calculation was done on backbone atoms of the protein pocket and the protein pocket was defined by atoms within 10 Å of ligand N3 in protein 6LU7 (Figure 2). A cut-off of 2.2 Å was used to create clusters. It was chosen in order to obtain 90% of the trajectory in the first 10 clusters. Finally centroïds of these 10 clusters were picked as inputs for our ensemble docking protocol (Figure 3).

The preparation of the protein includes the critical step of protonating the crystallographic structure. This influences the orientation and reactivity of residues such as histidines, glutamines and asparagines. As the catalytic dyad of the 3CLPro protein includes a histidine residue, this step is more important than ever. In order to be reactive, the catalytic cysteine needs to be able to go to its thiolate form (see Annex 3 for the full mechanism). To do so, the proton of the thiol is engaged in a hydrogen bond with the ε nitrogen of the catalytic histidine, meaning that its proton has to be on the δ nitrogen. Other residues with ambiguous protonation or topology were left as is from the molecular dynamics run after checking that they all made sense given the local environment of the residues.

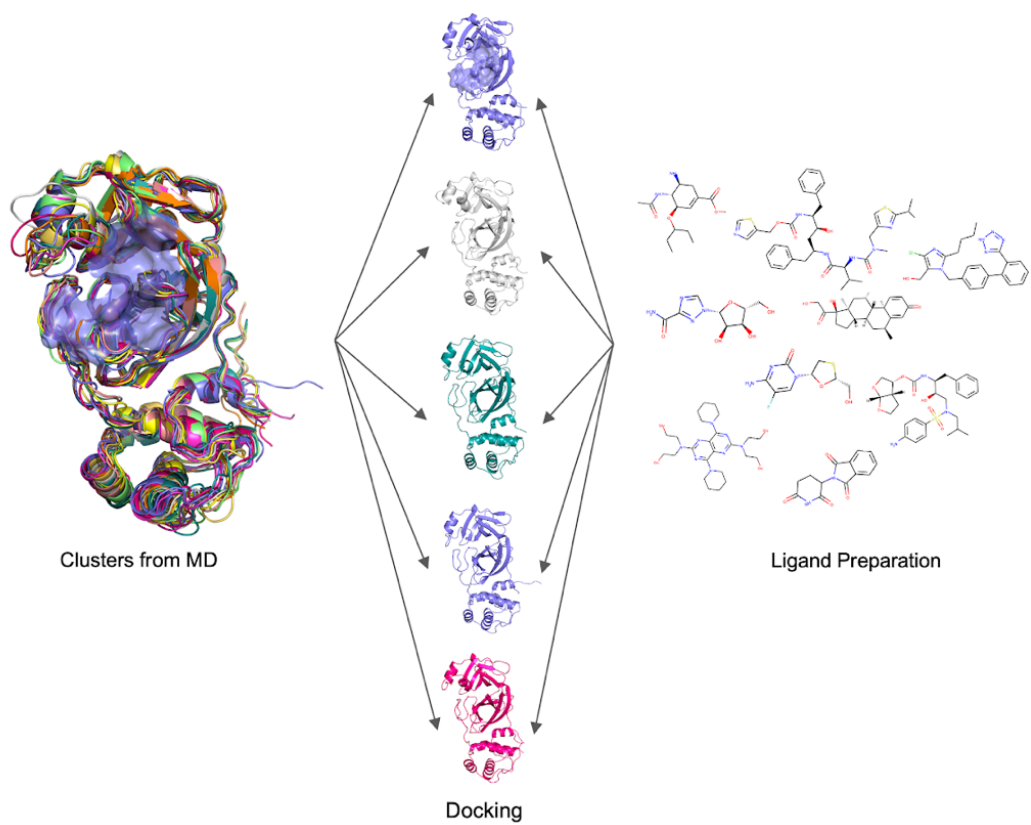

*Figure 4: Ensemble docking strategy aims at simulating the flexibility of a protein while reducing the cost of computing. Mainly used with snapshots extracted from a trajectory, these snapshots are cluster representatives of the most encountered conformations. On the left of the picture, alignment of all representatives. On the right, the in-house database of FDA-approved drugs. The database is then docked on each structure.*

## Docking and Screening

We have prepared a database comprising FDA approved drugs and prodrugs libraries[37] as well as known HIV and HCV proteases inhibitors[38] that were part of clinical trials. The database was filtered using criterias like a range of molecular weight from 250 to 900 g.mol$^{-1}$, typically 1-14 H-bond acceptors, 0-8 H-bond donors, and estimated logP values between 0 and 8. After filtering, the database contained about 1400 drugs already in use to cure various pathologies (antibiotics, antivirals, anti-inflammatory, anticancer, …). Prodrugs (compounds for which the administered molecule goes through a chemical change) or molecules without an oral administration route (such as the active ingredients in topical creams) were not removed from the database but rather labeled after the screening. Molecules protonation states were enumerated with chemaxon tools[46] and 3D generation was performed with RDkit[47]. Only the first three protonation states (according to chemaxon) were kept as they often represent the majority of the distribution for any given molecule. Charged protonation states were removed from the dataset as ranking can only be done between molecules of the same charge.

Autodock vina[39] was then used to generate 10 poses for each protonation state on each protein cluster centroïd (vina parameters : exhaustiveness 32, default parameters, rigid protein).

Non-specific interacting poses were filtered out using the contacts they made or did not make with specific residues in the binding pocket. We evaluated in greater detail the molecule's poses that interact with Cys 145, Ser 144 and G143 for their role in the catalytic process. We also chose Asn 142, Glu 166, Met 165 and Gln 189 due to their high prevalence in the energetic contribution of numerous inhibitors, as was said before.

Docking poses that passed our custom interactions filter were then rescored with our affinity predictor. The protein-ligand complex structures were prepared with the Amber16 tool set[40]. Proteins, ligands and water were represented by ff14SB[41], GAFF1.81[42] and TIP3P[43] models, respectively. The free energy of binding was then evaluated according to equation 1. Our prediction of binding free energy takes only a few minutes per compound, each running on 2 cpu cores (c5.large AWS EC2). Our docking/scoring protocol is automated and scaled on a dedicated, secured cloud. The docking and screening procedure of the 1400 molecules was completed in 30 minutes wall time, producing 2100+ scores.

# Results

The first 10 compounds with the highest affinity are ranked in Table 1. Their $\Delta G_{calc}$ range from -52.2 to -43.1 kcal/mol. Their placement in the $\Delta G_{calc}$ distribution can be found in the supplementary information (Annex 1). All $\Delta G_{calc}$ values along with the top 100 molecules can be found in the supplementary information as well (Annex 2). We decided to present only the first 10 molecules here, but molecules in the top 100 might be interesting to investigate.

*Table 1: top 10 best scored screened molecules.*

| rank | Commercial name and label | 2D Structure | rank | Commercial name and label | 2D Structure |
|---|---|---|---|---|---|
| 1 | Telinavir (HIV protease inhibitor, disc in phase II) | 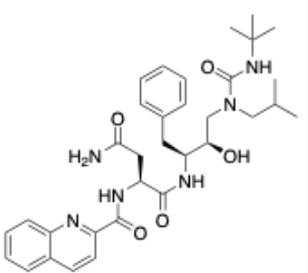 | 2 | Brecanavir (HIV protease inhibitor, disc in phase II) | 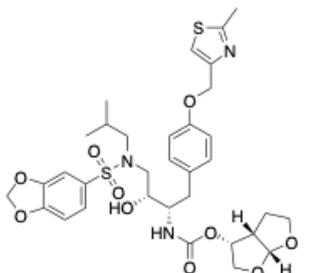 |
| 3 | Famotidine (FDA approved, orally available) | 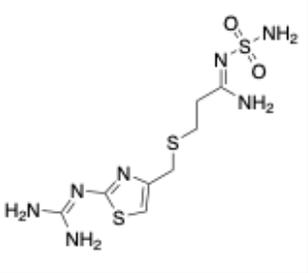 | 4 | Protokylol (FDA approved, orally available) | 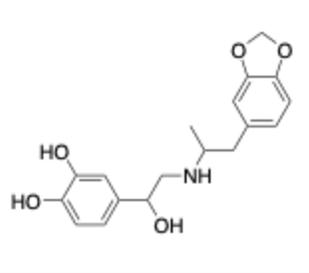 |
| 5 | Tenofovir alafenamide (FDA approved, prodrug) | 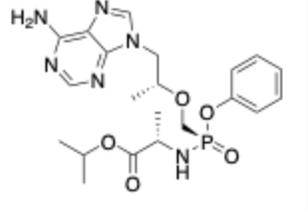 | 6 | Telaprevir (FDA approved, orally available) | 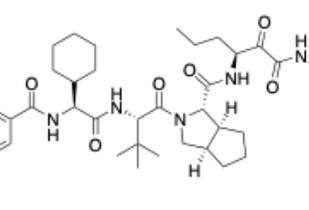 |
| 7 | Lasinavir (HIV protease inhibitor, disc in phase I) | 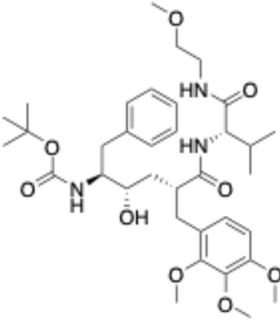 | 8 | Empagliflozin (FDA approved, orally available) | 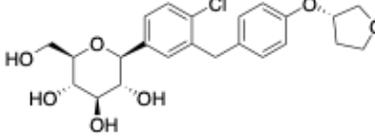 |
| 9 | Bacampicillin (FDA approved, prodrug) | 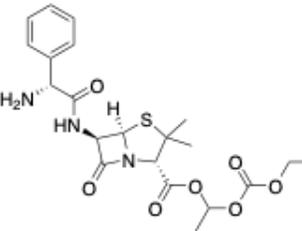 | 10 | Entecavir (FDA approved, orally available) | 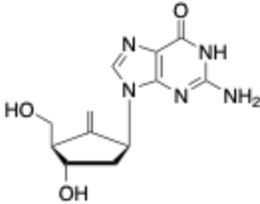 |

In the top 10, two are prodrugs - their active species are metabolites - that are unlikely to remain under this chemical form in a living organism and therefore to interact with the protein under these conditions. Still, they do offer insight on pharmacophores that are capable of binding and blocking the activity of 3CLPro.

Three HIV-antiviral drugs (Telinavir, Brecanavir and Lasinavir) have a strong predicted affinity with 3CLPro. These results are coherent with previous clinical trials of HIV-antivirals in SARS-CoV-1 disease (2003) where HIV-antiviral drugs have been proven to be active against the main protease[44].

In particular, Telinavir and Brecanavir exhibit very interesting binding poses (Figure 5 and Figure 6) filling the four 3CLPro sub-pockets. Unfortunately, no IC50 against SARS-CoV-2 3CLPro has been reported in the literature for HIV-protease inhibitors.

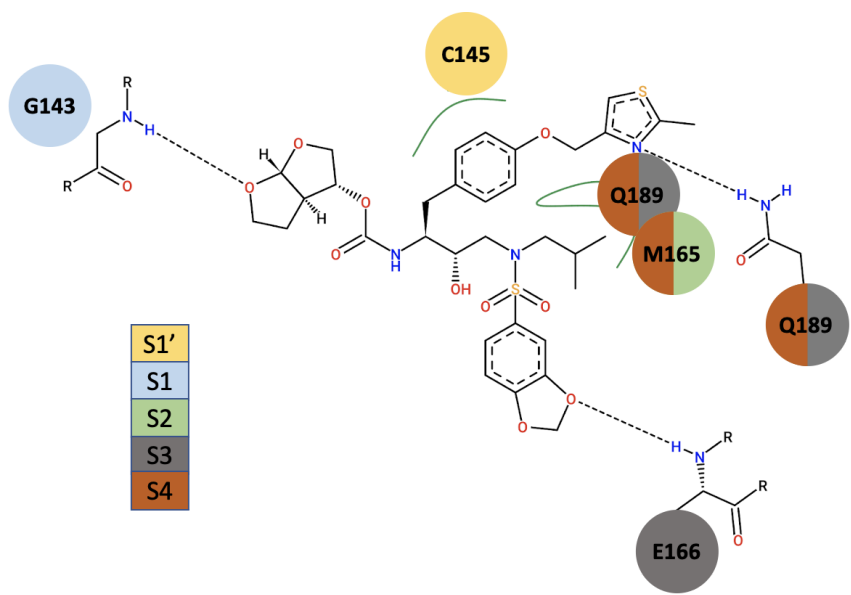

*Figure 5: 2D interaction map of brecanavir in 3CLPro. Hydrophobic interactions between 3CLPro and the molecule are represented in green solid lines and H-bond are represented by dotted lines. The color scale is used to label the 3CLPro pocket where the interaction is located.*

Famotidine has been measured inactive on 3CLPro[45] and is therefore, most likely, a false positive.

Finally, our screening highlighted other drugs (protokylol, empagliflozin, entecavir) which are not suspected to have activity against SARS-CoV-2 3CLPro and are not yet in clinical trials. Experimental measures will be needed to validate their antiviral activity against SARS-CoV-2.

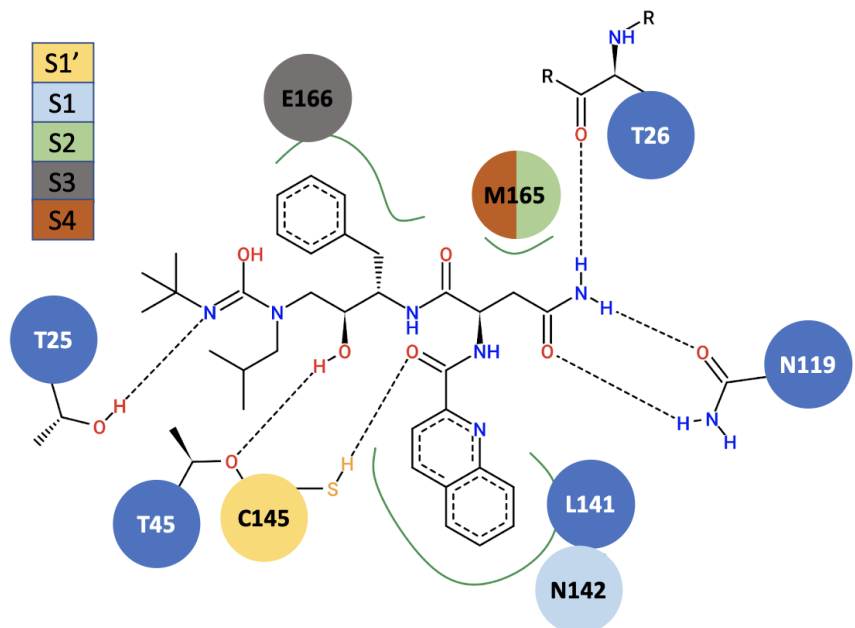

*Figure 6: 2D interaction map of telinavir in 3CLPro. Hydrophobic interactions between 3CLPro and the molecule are represented in green solid lines and H-bond are represented by dotted lines. The color scale is used to label the 3CLPro pocket where the interaction is located.*

Other HIV-antiviral drugs (Indinavir, Saquinavir, Tipranavir, Amprenavir, Darunavir, Palinavir, Nelfinavir and Mozenavir) are well-ranked (23rd, 28th, 34th, 40st, 55th, 68th, 96th and 100th). Vilazodone and dipyridamole, which are both measured as active against 3CLPro, are also well-ranked (42nd and 45th) in our screening.

## Conclusion

In this study, we leveraged Aqemia's absolute binding free energy calculation based on the molecular density functional theory to find known drugs with potential inhibition effects on 3CLPro. It is a first step toward the repositioning of available drugs in the treatment of COVID-19. Some of the highlighted drugs have already proven to be effective against SARS-CoV-2, some of these both in-vitro and clinically like Telaprevir, dipyridamole and vilazodone. We highlighted some drugs that are not in clinical trials yet and may help in the treatment of COVID-19. These *in silico* results need *in vitro* confirmation.

## Acknowledgment

This work was supported by Sanofi.

## About Aqemia

Aqemia is a deeptech startup in drug design. Our ambition is to discover better and more innovative therapeutic molecules faster. Better molecules because our physics-based technology has unparalleled precision. More innovative molecules because we don't rely on past data which allow us to explore a more diverse chemical space far from drugs on the market.  Faster because our precision requires less experiments.

Aqemia leverages a unique technology based on 8 years of research in quantum and statistical mechanics at Oxford (UK), Cambridge (UK) and École Normale Supérieure (Paris). Our algorithms are able to compute the affinity between molecules as precisely as traditional experiments and 10 000x faster than competition. This technology guides our AI towards the molecule that has maximum affinity for the therapeutic target and validates other biological and physico-chemical properties.

## References

(1) Official WHO count of the COVID-19 pandemic https://covid19.who.int/table (accessed 2020 -12 -14).
(2) Freedman, D. H. Hunting for New Drugs with AI. *Nature* **2019**, *576* (7787), S49–S53. https://doi.org/10.1038/d41586-019-03846-0.
(3) Jarada, T. N.; Rokne, J. G.; Alhajj, R. A Review of Computational Drug Repositioning:

(4) Xue, H.; Li, J.; Xie, H.; Wang, Y. Review of Drug Repositioning Approaches and Resources. *Int. J. Biol. Sci.* **2018**, *14* (10), 1232–1244. https://doi.org/10.7150/ijbs.24612.

(5) Luukkonen, S.; Belloni, L.; Borgis, D.; Levesque, M. High-Throughput Free Energies and Water Maps for Drug Discovery by Molecular Density Functional Theory. *ArXiv180603118 Phys.* **2018**.

(6) Wu, A.; Peng, Y.; Huang, B.; Ding, X.; Wang, X.; Niu, P.; Meng, J.; Zhu, Z.; Zhang, Z.; Wang, J.; Sheng, J.; Quan, L.; Xia, Z.; Tan, W.; Cheng, G.; Jiang, T. Genome Composition and Divergence of the Novel Coronavirus (2019-NCoV) Originating in China. *Cell Host Microbe* **2020**, *27* (3), 325–328. https://doi.org/10.1016/j.chom.2020.02.001.

(7) V'kovski, P.; Kratzel, A.; Steiner, S.; Stalder, H.; Thiel, V. Coronavirus Biology and Replication: Implications for SARS-CoV-2. *Nat. Rev. Microbiol.* **2020**, 1–16. https://doi.org/10.1038/s41579-020-00468-6.

(8) Ratia, K.; Saikatendu, K. S.; Santarsiero, B. D.; Barretto, N.; Baker, S. C.; Stevens, R. C.; Mesecar, A. D. Severe Acute Respiratory Syndrome Coronavirus Papain-like Protease: Structure of a Viral Deubiquitinating Enzyme. *Proc. Natl. Acad. Sci.* **2006**, *103* (15), 5717–5722. https://doi.org/10.1073/pnas.0510851103.

(9) Wu, F.; Zhao, S.; Yu, B.; Chen, Y.-M.; Wang, W.; Song, Z.-G.; Hu, Y.; Tao, Z.-W.; Tian, J.-H.; Pei, Y.-Y.; Yuan, M.-L.; Zhang, Y.-L.; Dai, F.-H.; Liu, Y.; Wang, Q.-M.; Zheng, J.-J.; Xu, L.; Holmes, E. C.; Zhang, Y.-Z. A New Coronavirus Associated with Human Respiratory Disease in China. *Nature* **2020**, *579* (7798), 265–269. https://doi.org/10.1038/s41586-020-2008-3.

(10) Zhou, P.; Yang, X.-L.; Wang, X.-G.; Hu, B.; Zhang, L.; Zhang, W.; Si, H.-R.; Zhu, Y.; Li, B.; Huang, C.-L.; Chen, H.-D.; Chen, J.; Luo, Y.; Guo, H.; Jiang, R.-D.; Liu, M.-Q.; Chen, Y.; Shen, X.-R.; Wang, X.; Zheng, X.-S.; Zhao, K.; Chen, Q.-J.; Deng, F.; Liu, L.-L.; Yan, B.; Zhan, F.-X.; Wang, Y.-Y.; Xiao, G.-F.; Shi, Z.-L. A Pneumonia Outbreak Associated with a New Coronavirus of Probable Bat Origin. *Nature* **2020**, *579* (7798), 270–273. https://doi.org/10.1038/s41586-020-2012-7.

(11) Anand, K.; Ziebuhr, J.; Wadhwani, P.; Mesters, J. R.; Hilgenfeld, R. Coronavirus Main Proteinase (3CLpro) Structure: Basis for Design of Anti-SARS Drugs. *Science* **2003**, *300* (5626), 1763–1767. https://doi.org/10.1126/science.1085658.

(12) Pillaiyar, T.; Manickam, M.; Namasivayam, V.; Hayashi, Y.; Jung, S.-H. An Overview of Severe Acute Respiratory Syndrome–Coronavirus (SARS-CoV) 3CL Protease Inhibitors: Peptidomimetics and Small Molecule Chemotherapy. *J. Med. Chem.* **2016**, *59* (14), 6595–6628. https://doi.org/10.1021/acs.jmedchem.5b01461.

(13) Agbowuro, A. A.; Huston, W. M.; Gamble, A. B.; Tyndall, J. D. A. Proteases and Protease Inhibitors in Infectious Diseases. *Med. Res. Rev.* **2018**, *38* (4), 1295–1331. https://doi.org/10.1002/med.21475.

(14) Ghosh, A. K.; Xi, K.; Ratia, K.; Santarsiero, B. D.; Fu, W.; Harcourt, B. H.; Rota, P. A.; Baker, S. C.; Johnson, M. E.; Mesecar, A. D. Design and Synthesis of Peptidomimetic Severe Acute Respiratory Syndrome Chymotrypsin-like Protease Inhibitors. *J. Med. Chem.* **2005**, *48* (22), 6767–6771. https://doi.org/10.1021/jm050548m.

(15) Kumar, V.; Tan, K.-P.; Wang, Y.-M.; Lin, S.-W.; Liang, P.-H. Identification, Synthesis and Evaluation of SARS-CoV and MERS-CoV 3C-like Protease Inhibitors. *Bioorg. Med. Chem.* **2016**, *24* (13), 3035–3042. https://doi.org/10.1016/j.bmc.2016.05.013.

Strategies, Approaches, Opportunities, Challenges, and Directions. *J. Cheminformatics* **2020**, *12* (1), 46. https://doi.org/10.1186/s13321-020-00450-7.

(16) Świderek, K.; Moliner, V. Revealing the Molecular Mechanisms of Proteolysis of SARS-CoV-2 Mpro by QM/MM Computational Methods. *Chem. Sci.* **2020**, *11* (39), 10626–10630. https://doi.org/10.1039/D0SC02823A.

(17) Goyal, B.; Goyal, D. Targeting the Dimerization of the Main Protease of Coronaviruses: A Potential Broad-Spectrum Therapeutic Strategy. *ACS Comb. Sci.* **2020**, *22* (6), 297–305. https://doi.org/10.1021/acscombsci.0c00058.

(18) S. Hosseini-Zare, M.; Thilagavathi, R.; Selvam, C. Targeting Severe Acute Respiratory Syndrome-Coronavirus (SARS-CoV-1) with Structurally Diverse Inhibitors: A Comprehensive Review. *RSC Adv.* **2020**, *10* (47), 28287–28299. https://doi.org/10.1039/D0RA04395H.

(19) Yang, H.; Xie, W.; Xue, X.; Yang, K.; Ma, J.; Liang, W.; Zhao, Q.; Zhou, Z.; Pei, D.; Ziebuhr, J.; Hilgenfeld, R.; Yuen, K. Y.; Wong, L.; Gao, G.; Chen, S.; Chen, Z.; Ma, D.; Bartlam, M.; Rao, Z. Design of Wide-Spectrum Inhibitors Targeting Coronavirus Main Proteases. *PLOS Biol.* **2005**, *3* (10), e324. https://doi.org/10.1371/journal.pbio.0030324.

(20) Ullrich, S.; Nitsche, C. The SARS-CoV-2 Main Protease as Drug Target. *Bioorg. Med. Chem. Lett.* **2020**, *30* (17), 127377. https://doi.org/10.1016/j.bmcl.2020.127377.

(21) Zhang, L.; Lin, D.; Sun, X.; Curth, U.; Drosten, C.; Sauerhering, L.; Becker, S.; Rox, K.; Hilgenfeld, R. Crystal Structure of SARS-CoV-2 Main Protease Provides a Basis for Design of Improved α-Ketoamide Inhibitors. *Science* **2020**, *368* (6489), 409–412. https://doi.org/10.1126/science.abb3405.

(22) Zhang, L.; Lin, D.; Kusov, Y.; Nian, Y.; Ma, Q.; Wang, J.; von Brunn, A.; Leyssen, P.; Lanko, K.; Neyts, J.; de Wilde, A.; Snijder, E. J.; Liu, H.; Hilgenfeld, R. α-Ketoamides as Broad-Spectrum Inhibitors of Coronavirus and Enterovirus Replication: Structure-Based Design, Synthesis, and Activity Assessment. *J. Med. Chem.* **2020**, *63* (9), 4562–4578. https://doi.org/10.1021/acs.jmedchem.9b01828.

(23) Hilgenfeld, R. From SARS to MERS: Crystallographic Studies on Coronaviral Proteases Enable Antiviral Drug Design. *FEBS J.* **2014**, *281* (18), 4085–4096. https://doi.org/10.1111/febs.12936.

(24) Khan, S. A.; Zia, K.; Ashraf, S.; Uddin, R.; Ul-Haq, Z. Identification of Chymotrypsin-like Protease Inhibitors of SARS-CoV-2 via Integrated Computational Approach. *J. Biomol. Struct. Dyn.* **2020**, *0* (0), 1–10. https://doi.org/10.1080/07391102.2020.1751298.

(25) Das, P.; Majumder, R.; Mandal, M.; Basak, P. In-Silico Approach for Identification of Effective and Stable Inhibitors for COVID-19 Main Protease (Mpro) from Flavonoid Based Phytochemical Constituents of Calendula Officinalis. *J. Biomol. Struct. Dyn.* **2020**, *0* (0), 1–16. https://doi.org/10.1080/07391102.2020.1796799.

(26) Keretsu, S.; Bhujbal, S. P.; Cho, S. J. Rational Approach toward COVID-19 Main Protease Inhibitors via Molecular Docking, Molecular Dynamics Simulation and Free Energy Calculation. *Sci. Rep.* **2020**, *10* (1), 17716. https://doi.org/10.1038/s41598-020-74468-0.

(27) Duy Nguyen, D.; Gao, K.; Chen, J.; Wang, R.; Wei, G.-W. Unveiling the Molecular Mechanism of SARS-CoV-2 Main Protease Inhibition from 137 Crystal Structures Using Algebraic Topology and Deep Learning. *Chem. Sci.* **2020**, *11* (44), 12036–12046. https://doi.org/10.1039/D0SC04641H.

(28) Singh, J.; Petter, R. C.; Baillie, T. A.; Whitty, A. The Resurgence of Covalent Drugs. *Nat. Rev. Drug Discov.* **2011**, *10* (4), 307–317. https://doi.org/10.1038/nrd3410.

(29) Sisay, M. 3CLpro Inhibitors as a Potential Therapeutic Option for COVID-19: Available Evidence and Ongoing Clinical Trials. *Pharmacol. Res.* **2020**, *156*, 104779.

https://doi.org/10.1016/j.phrs.2020.104779.
(30) Cusinato, J.; Cau, Y.; Calvani, A. M.; Mori, M. Repurposing Drugs for the Management of COVID-19. *Expert Opin. Ther. Pat.* **2020**, *0* (ja), null. https://doi.org/10.1080/13543776.2021.1861248.
(31) Drayman, N.; Jones, K. A.; Azizi, S.-A.; Froggatt, H. M.; Tan, K.; Maltseva, N. I.; Chen, S.; Nicolaescu, V.; Dvorkin, S.; Furlong, K.; Kathayat, R. S.; Firpo, M. R.; Mastrodomenico, V.; Bruce, E. A.; Schmidt, M. M.; Jedrzejczak, R.; Muñoz-Alía, M. Á.; Schuster, B.; Nair, V.; Botten, J. W.; Brooke, C. B.; Baker, S. C.; Mounce, B. C.; Heaton, N. S.; Dickinson, B. C.; Jaochimiak, A.; Randall, G.; Tay, S. Drug Repurposing Screen Identifies Masitinib as a 3CLpro Inhibitor That Blocks Replication of SARS-CoV-2 in Vitro. *bioRxiv* **2020**. https://doi.org/10.1101/2020.08.31.274639.
(32) Mittal, L.; Kumari, A.; Srivastava, M.; Singh, M.; Asthana, S. Identification of Potential Molecules against COVID-19 Main Protease through Structure-Guided Virtual Screening Approach. *J. Biomol. Struct. Dyn.* **2020**, *0* (0), 1–19. https://doi.org/10.1080/07391102.2020.1768151.
(34) Liu, X.; Li, Z.; Liu, S.; Sun, J.; Chen, Z.; Jiang, M.; Zhang, Q.; Wei, Y.; Wang, X.; Huang, Y.-Y.; Shi, Y.; Xu, Y.; Xian, H.; Bai, F.; Ou, C.; Xiong, B.; Lew, A. M.; Cui, J.; Fang, R.; Huang, H.; Zhao, J.; Hong, X.; Zhang, Y.; Zhou, F.; Luo, H.-B. Potential Therapeutic Effects of Dipyridamole in the Severely Ill Patients with COVID-19. *Acta Pharm. Sin. B* **2020**. https://doi.org/10.1016/j.apsb.2020.04.008.
(35) D. E. Shaw Research. Molecular Dynamics Simulations Related to SARS-CoV-2. **2020**, *http://www.deshawresearch.com/resources_sarscov2.html*.
(36) Daura, X.; Gademann, K.; Jaun, B.; Seebach, D.; Gunsteren, W. F. van; Mark, A. E. Peptide Folding: When Simulation Meets Experiment. *Angew. Chem. Int. Ed.* **1999**, *38* (1–2), 236–240. https://doi.org/10.1002/(SICI)1521-3773(19990115)38:1/2<236::AID-ANIE236>3.0.CO;2-M.
(37) Center for Drug Evaluation and. Drugs@FDA Data Files https://www.fda.gov/drugs/drug-approvals-and-databases/drugsfda-data-files (accessed 2020 -02 -29).
(38) Wishart, D. S.; Feunang, Y. D.; Guo, A. C.; Lo, E. J.; Marcu, A.; Grant, J. R.; Sajed, T.; Johnson, D.; Li, C.; Sayeeda, Z.; Assempour, N.; Iynkkaran, I.; Liu, Y.; Maciejewski, A.; Gale, N.; Wilson, A.; Chin, L.; Cummings, R.; Le, D.; Pon, A.; Knox, C.; Wilson, M. DrugBank 5.0: A Major Update to the DrugBank Database for 2018. *Nucleic Acids Res.* **2018**, *46* (D1), D1074–D1082. https://doi.org/10.1093/nar/gkx1037.
(39) Trott, O.; Olson, A. J. AutoDock Vina: Improving the Speed and Accuracy of Docking with a New Scoring Function, Efficient Optimization and Multithreading. *J. Comput. Chem.* **2010**, *31* (2), 455–461. https://doi.org/10.1002/jcc.21334.
(40) D.A. Case, H.M. Aktulga, K. Belfon, I.Y. Ben-Shalom, S.R. Brozell, D.S. Cerutti, T.E. Cheatham, III, V.W.D. Cruzeiro, T.A. Darden, R.E. Duke, G. Giambasu, M.K. Gilson, H. Gohlke, A.W. Goetz, R. Harris, S. Izadi, S.A. Izmailov, C. Jin, K. Kasavajhala, M.C. Kaymak, E. King, A. Kovalenko, T. Kurtzman, T.S. Lee, S. LeGrand, P. Li, C. Lin, J. Liu, T. Luchko, R. Luo, M. Machado, V. Man, M. Manathunga, K.M. Merz, Y. Miao, O. Mikhailovskii, G. Monard, H. Nguyen, K.A. O'Hearn, A. Onufriev, F. Pan, S. Pantano, R. Qi, A. Rahnamoun, D.R. Roe, A. Roitberg, C. Sagui, S. Schott-Verdugo, J. Shen, C.L. Simmerling, N.R. Skrynnikov, J. Smith, J.

Swails, R.C. Walker, J. Wang, H. Wei, R.M. Wolf, X. Wu, Y. Xue, D.M. York, S. Zhao, and P.A. Kollman (2021), Amber 2021, University of California, San Francisco.

(41) Maier, J. A.; Martinez, C.; Kasavajhala, K.; Wickstrom, L.; Hauser, K. E.; Simmerling, C. Ff14SB: Improving the Accuracy of Protein Side Chain and Backbone Parameters from Ff99SB. *J. Chem. Theory Comput.* **2015**, *11* (8), 3696–3713. https://doi.org/10.1021/acs.jctc.5b00255.

(42) Wang, J.; Wolf, R. M.; Caldwell, J. W.; Kollman, P. A.; Case, D. A. Development and Testing of a General Amber Force Field. *J. Comput. Chem.* **2004**, *25* (9), 1157–1174. https://doi.org/10.1002/jcc.20035.

(43) Jorgensen, W. L.; Chandrasekhar, J.; Madura, J. D.; Impey, R. W.; Klein, M. L. Comparison of Simple Potential Functions for Simulating Liquid Water. *J. Chem. Phys.* **1983**, *79* (2), 926–935. https://doi.org/10.1063/1.445869.

(44) Savarino, A. Expanding the Frontiers of Existing Antiviral Drugs: Possible Effects of HIV-1 Protease Inhibitors against SARS and Avian Influenza. *J. Clin. Virol.* **2005**, *34* (3), 170–178. https://doi.org/10.1016/j.jcv.2005.03.005.

(46) Loffredo, M.; Lucero, H.; Chen, D.-Y.; O'Connell, A.; Bergqvist, S.; Munawar, A.; Bandara, A.; De Graef, S.; Weeks, S. D.; Douam, F.; Saeed, M.; Munawar, A. H. The In-Vitro Effect of Famotidine on Sars-Cov-2 Proteases and Virus Replication. *Sci. Rep.* **2021**, *11* (1), 5433. https://doi.org/10.1038/s41598-021-84782-w.

(47) Contini, A. Virtual Screening of an FDA Approved Drugs Database on Two COVID-19 Coronavirus Proteins. **2020**. https://doi.org/10.26434/chemrxiv.11847381.v1.

(46) cxcalc dominanttautomerdistribution was/were used for protonation states generation, cxcalc dominanttautomerdistribution, ChemAxon (https://www.chemaxon.com).

(47) RDKit: Open-source cheminformatics; http://www.rdkit.org

**Supplementary informations**

Annex 1 : distribution of $\Delta G_{calc}$ (kcal/mol) on the ligand database (one pose per compound).

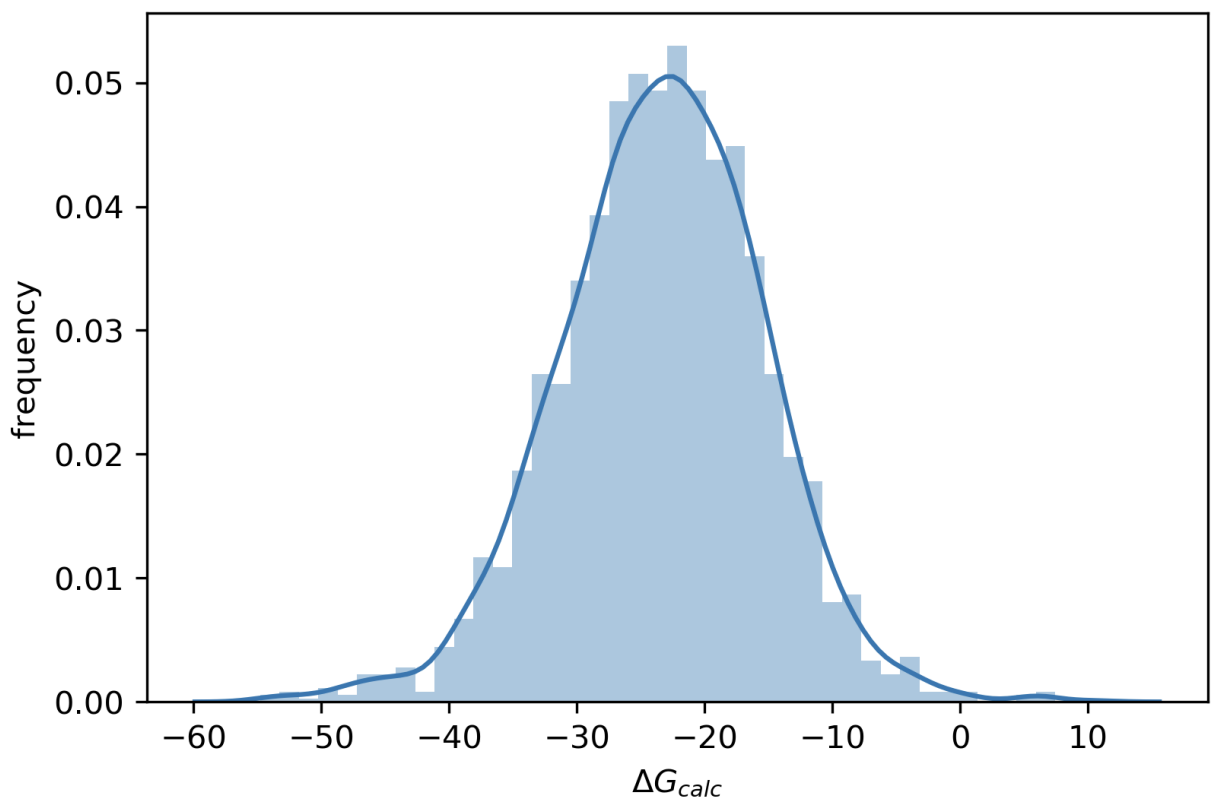

Annex 2 : top 100 best scored molecules coming from our screening.

| Rank | Name | $\Delta G_{calc}$ (kcal/mol) | Rank | Name | $\Delta G_{calc}$ (kcal/mol) |
|---|---|---|---|---|---|
| 1 | Telinavir | -52.2 | 51 | Dorzolamide | -38.2 |
| 2 | Brecanavir | -50.7 | 52 | Xamoterol | -38.0 |
| 3 | Famotidine | -49.1 | 53 | Valganciclovir | -37.9 |
| 4 | Protokylol | -49.0 | 54 | Clocortolone | -37.9 |
| 5 | Tenofovir Alafenamide | -47.1 | 55 | Darunavir | -37.8 |
| 6 | Telaprevir | -46.9 | 56 | Tucatinib | -37.8 |
| 7 | Lasinavir | -45.7 | 57 | Macitentan | -37.8 |
| 8 | Empagliflozin | -44.9 | 58 | Mirabegron | -37.8 |
| 9 | Bacampicillin | -44.8 | 59 | Maraviroc | -37.5 |
| 10 | Entecavir | -43.1 | 60 | Latanoprost | -37.4 |
| 11 | Fenoterol | -42.8 | 61 | Desonide | -37.3 |
| 12 | Curcumin | -42.1 | 62 | Doxycycline | -37.3 |
| 13 | Hydrocortisone Butyrate | -42.0 | 63 | Telithromycin | -37.2 |
| 14 | Dexamethasone | -41.9 | 64 | Triamcinolone | -37.2 |
| 15 | Darolutamide | -41.8 | 65 | Sildenafil | -37.2 |
| 16 | Ranolazine | -41.8 | 66 | Labetalol | -37.2 |
| 17 | Tezacaftor | -41.5 | 67 | Iobitridol | -37.1 |
| 18 | Arzoxifene | -40.9 | 68 | Palinavir | -36.9 |
| 19 | Triamcinolone Acetonide | -40.8 | 69 | Lurasidone | -36.8 |
| 20 | Betamethasone | -40.6 | 70 | Indacaterol | -36.8 |
| 21 | Onapristone | -40.2 | 71 | Regadenoson | -36.8 |
| 22 | Ticagrelor | -40.2 | 72 | Copanlisib | -36.7 |
| 23 | Indinavir | -40.2 | 73 | Almitrine | -36.7 |
| 24 | Canagliflozin | -40.1 | 74 | Ioxilan | -36.7 |
| 25 | Cianidanol | -40.1 | 75 | Olodaterol | -36.6 |
| 26 | Telotristat Ethyl | -40.0 | 76 | Gemeprost | -36.6 |
| 27 | Metrizamide | -39.9 | 77 | Fenticonazole | -36.5 |
| 28 | Saquinavir | -39.9 | 78 | Chlortetracycline | -36.5 |
| 29 | Elvitegravir | -39.8 | 79 | Sorafenib | -36.5 |
| 30 | Brigatinib | -39.6 | 80 | Naltrexone | -36.5 |
| 31 | Cobicistat | -39.6 | 81 | Donepezil | -36.5 |
| 32 | Valaciclovir | -39.6 | 82 | Dabrafenib | -36.5 |
| 33 | Fedratinib | -39.3 | 83 | Glafenine | -36.4 |
| 34 | Tipranavir | -39.2 | 84 | Etoposide | -36.3 |
| 35 | Hexobendine | -39.2 | 85 | Lorpiprazole | -36.3 |
| 36 | Trimetrexate | -39.2 | 86 | Oxatomide | -36.3 |

| 37 | Tetracycline | -39.0 | 87 | Betaxolol | -36.2 |
| --- | --- | --- | --- | --- | --- |
| 38 | Vilanterol | -39.0 | 88 | Cabozantinib | -36.2 |
| 39 | Hydrocortisone | -39.0 | 89 | Deflazacort | -36.2 |
| 40 | Amprenavir | -38.9 | 90 | Loperamide | -36.1 |
| 41 | Carvedilol | -38.9 | 91 | Prednisone | -36.1 |
| 42 | Vilazodone | -38.8 | 92 | Dasabuvir | -36.0 |
| 43 | Dobutamine | -38.4 | 93 | Lercanidipine | -36.0 |
| 44 | Rifaximin | -38.4 | 94 | Delafloxacin | -36.0 |
| 45 | Dipyridamole | -38.3 | 95 | Salmeterol | -35.9 |
| 46 | Budesonide | -38.3 | 96 | Nelfinavir | -35.9 |
| 47 | Arbutamine | -38.3 | 97 | Verapamil | -35.9 |
| 48 | Fluprednisolone | -38.3 | 98 | Trazodone | -35.7 |
| 49 | Ximelagatran | -38.2 | 99 | Perphenazine | -35.5 |
| 50 | Prednisolone | -38.2 | 100 | Mozenavir | -35.5 |

Annex 3: Mechanism of 3CLPro hydrolysis process[16]

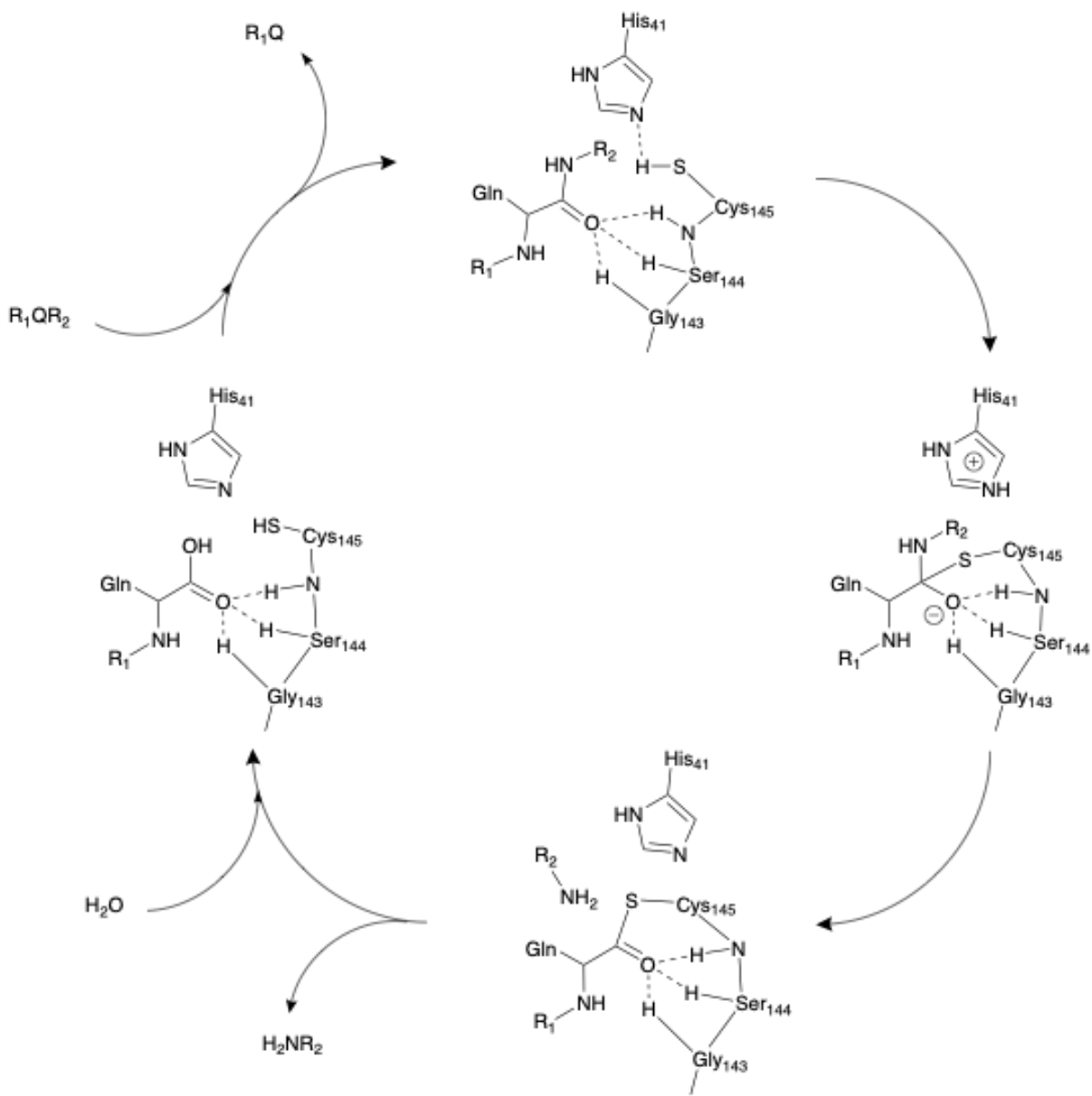

Annex 4: Top 10 FDA-approved, no-pro drug, orally available

| Rank (global) | Name |
|---|---|
| 3 | Famotidine |
| 4 | Protokylol |
| 6 | Telaprevir |
| 8 | Empagliflozin |
| 10 | Entecavir |
| 14 | Dexamethasone |
| 15 | Darolutamide |
| 16 | Ranolazine |
| 17 | Tezacaftor |
| 18 | Arzoxifene |

Annex 5: Top HIV or HCV antiproteases

| Rank (global) | Name |
|---|---|
| 1 | Telinavir |
| 2 | Brecanavir |
| 6 | Telaprevir |
| 7 | Lasinavir |
| 28 | Saquinavir |
| 55 | Darunavir |
| 68 | Palinavir |
| 100 | Mozenavir |